\documentclass[aps,nofootinbib,preprintnumbers,citeautoscript]{revtex4}
\usepackage{amsmath,amssymb}
\usepackage{enumerate}
\usepackage{graphicx}
\date{\today}
\begin{document}
\title{Influence of Dzyaloshinshkii-Moriya interaction on quantum correlations in two qubit Werner states and MEMS}

\author{Kapil K. Sharma$^\ast$ and  S. N. Pandey$^\dagger$  \\
\textit{Department of Physics, Motilal Nehru National Institute of Technology,\\ Allahabad 211004, India.} \\
E-mail: $^\ast$scienceglobal@gmail.com, $^\dagger$snp@mnnit.ac.in}

\begin{abstract}
In this paper we study the influence of Dzyaloshinskii-Moriya (DM) interaction on quantum correlations in two qubit Werner states and maximally entangled mixed states (MEMS). We consider our system as a closed system of a qubits pair and one auxiliary qubit which interact with any one of the qubit of the pair through DM interaction. We show that DM interaction, taken along any direction (x or y or z), does not affect two qubit Werner states. On the other hand the MEMS are affected by x and z components of DM interaction and remain unaffected by the y component. Further, we find that the state (i.e probability amplitude) of auxiliary qubit do not affect the quantum correlations in both the states, only DM interaction strength influences the quantum correlations. So one can avoid the intention to prepare the specific state of auxiliary qubit to manipulate the quantum correlations in both the states. We mention here that avoiding the preparation of state can contribute to cost reduction in quantum information processing. We also observe the phenomenon of entanglement sudden death in the present study. 
\end{abstract}
\maketitle

%\newpage

%\tableofcontents
\section{Introduction}
\label{intro}
Entanglement \cite{EPR1935,Neilsen2000} and quantum discord \cite{discord} are the two aspects of quantum correlations and have the great potential in quantum information processing, which have many applications for future quantum technologies. The prominent role of entanglement have been investigated in teleportation, quantum secure cryptography, quantum games, quantum imaging, superdense coding and so on \cite{CBennet1933,AkEkert1991,qs,qi,super}
. The quantum correlations are very much fragile to the environmental interactions and they may loose their correlations. As a result they may lead to entanglement sudden death (ESD) for finite time \cite{YuEberly2004,YuEberly2009}. The phenomenon of ESD in quantum information is a subject of  investigation for practical quantum information processing. 

The two qubit system is the simplest bipartite system in quantum information that can be prepared either in pure or mixed bipartite quantum entangled states. Werner states are quantum states which are mixed two qubit entangled states \cite{werner}. Munro et al. have investigated another class of maximally entangled mixed states (MEMS) \cite{MEMS}, which have more entanglement than Werner states and these are experimentally verified also \cite{exMEMS}. Two qubit states have been realized in a variety of physical systems. So it is interesting to study the evolution of quantum correlations in physical systems under various interactions. In this paper we study the influence of Dzyaloshinshkii-Moriya (DM) interaction \cite{DMinteraction,DMmoriya,Tmoriya2_1960} on the dynamics of entanglement and quantum discord \cite{SLu} in two qubit mixed bipartite states (i.e. Werner states and MEMS). We consider a pair of qubits and one auxiliary qubit, which interact with any one of the qubits of the pair through DM interaction. DM interaction plays a major role in quantum information processing and there is a vast literature in which the various quantum spin chains embedded in optical lattices governed by different models (Heisenberg, Ising, frustrated etc.)  
 \cite{ultra_cold,optical_lattices} have been studied under DM interaction. DM interaction plays an important role to entangle, disentangle and amplifying the entanglement in quantum systems. But sometimes it may be the killer of entanglement in some quantum states. This has been shown by Zang et al. by considering two qubits A and B prepared in Bell states \cite{ZQiang1,ZQiang2,ZQiang3} and a third qubit C which interact via DM interaction with the qubit B. They have shown that DM interaction amplify and periodically kills the entanglement in two qubit Bell states. The entanglement manipulation can be done by controlling the state (i.e. Probability amplitude) of the third qubit (C) and DM interaction strength. Our observation tells that DM interaction is always not the killer of entanglement in quantum states as some of these remain unaffected by the same \cite{kk}. Recently, we also have shown that the state of  auxiliary qubit (i.e. Probability amplitude) do not influence the entanglement in two parameter qubit-qutrit states \cite{two_parameter}. In the present paper we show that the state of auxiliary qubit (i.e probability amplitude) does not come into picture to manipulate the quantum correlations in Werner states and MEMS; only DM interaction strength influences the quantum correlations. So in this case one can avoid to prepare the specific state of auxiliary qubit to manipulate the entanglement in both Werner states and MEMS. Avoiding the state preparation can help in cost reduction in quantum information processing. Recently the quantum networks have been established under DM interaction \cite{N_M_1,N_M_2,N_M_3}, so the present study may be useful in manipulating and controlling the quantum correlations in quantum networks under DM interaction.

The plan of the paper is as follows. In Sect. 2 we present the Hamiltonian of the system. In Sect. 3 we discuss the two qubit Werner states and MEMS. In Sect. 4 we study the reduce density matrix (RDM) and eigenvalue spectrum. We study the influence of  DM interaction on quantum correlations in Werner states and MEMS in Sect 5. Further, in Sect 6. we present our conclusion. We present the quantum discord in Appendix and mathematica code to calculate the quantum discord for X-structured density matrices.
\section{Hamiltonian of the system}
We consider a qubit (A)-qubit (B) pair and an auxiliary qubit (C) which interact with the qubit (B) of the pair through DM interaction. The Hamiltonian of the system can be framed as
\begin{equation}
H=H_{AB}+H_{BC}^{int}, \label{eq:17}
\end{equation}
where $H_{AB}$ is the Hamiltonian of qubit (A) and qubit (B) and $H_{BC}^{int}$ is the interaction Hamiltonian of qubit (B) and qubit (C). Here we consider uncoupled qubit (A) and qubit (B), so $H_{AB}$ is zero. Now the Hamiltonian becomes
\begin{eqnarray}
H=H_{BC}^{int}=\vec{D}.(\vec{\sigma_B} \times \vec{\sigma_C}),   \label{eq:18}
\end{eqnarray}
where $\vec{D}$ is DM interaction between qubit (B) and qubit (C), $\vec{\sigma_B}$ and $\vec{\sigma_C}$ are the Pauli vectors associated with qubit (B) and qubit (C) whose components are Pauli matrices. 
We assume that DM interaction exist along the x, y and z-directions. In this case the Hamiltonian can be simplified as
\begin{center}
\begin{eqnarray}
H_{x}=D_{x}.(\sigma_B^Y \otimes \sigma_C^Z-\sigma_B^Z \otimes \sigma_C^Y),    \label{eq:Hx} \\  
H_{y}=D_{y}.(\sigma_B^Z \otimes \sigma_C^X-\sigma_B^X \otimes \sigma_C^Z),    \label{eq:Hy} \\  
H_{z}=D_{z}.(\sigma_B^X \otimes \sigma_C^Y-\sigma_B^Y \otimes \sigma_C^X),    \label{eq:Hz}   
\end{eqnarray}
\end{center}
where $\sigma_B^X$, $\sigma_B^Y$ and $\sigma_B^Z$
are  Pauli matrices for qubit (B) and $\sigma_C^X$, $\sigma_C^Y$ and $\sigma_C^Z$ are Pauli matrices of qubit (C). The above Hamiltonians are the matrices having $4\times 4$ dimension and easy to diagonalize by using the method of eigendecomposition. The unitary time evolution operator is easily commuted  as 
\begin{eqnarray}
U(t)=e^{-i H t}, \label{eq:20}
\end{eqnarray}
 which is also a $4 \times 4$ matrix. This operator has been used to obtain the time evolution of the system.
\section{Two qubit Werner states and MEMS}
In this section we discuss the two qubit bipartite Werner states \cite{werner} and MEMS \cite{MEMS}. The Werner states in $2\times 2$ dimensions are widely used in quantum information and are invariant under unitary operators. The two qubit MEMS have been investigated by Munro et al., which are more entangled than Werner states \cite{MEMS}. The entanglement in these states can be calculated by using the concurrence. The concurrence $C$ in a density matrix 
$\rho$ is given by 
\begin{equation}
C(\rho)=max\{0,\lambda_{1}-\lambda_{2}-\lambda_{3}-\lambda_{4}\},
\end{equation}
where $\lambda_{i}$, $i=1,2,3,4$, are the square roots of the eigenvalues in decreasing order of $\rho \rho^{f}$, $\rho^{f}$ is the spin flip density matrix given as \\
\begin{equation}
\rho^{f}=(\sigma^{y}\otimes \sigma^{y}) \rho^{*}( \sigma^{y}\otimes \sigma^{y}). 
\end{equation}
Here $\rho^{*}$ is the complex conjugate of the density matrix $\rho$ and $\sigma^{y}$ is the Pauli Y matrix. The matrix $\rho \rho^{f}$ is used to calculate the concurrence. 
Two qubit Werner states can be written as 
\begin{equation}
\rho_{w}=\gamma |\psi^{-}\rangle \langle \psi^{-}|+(1-\gamma)\frac{I}{4}. \label{eq:1}
\end{equation}
where $|\psi^{-}\rangle$ is the singlet state and $I$ is the $4 \times 4$ identity matrix. The state is entangled with $\gamma>1/3$ and carry the concurrence as $C(\gamma)=(3\gamma -1)/2$. The MEMS are given below as 
\begin{center}
\begin{equation}
\rho_{MEMS}= \left[ \begin{array}{cccc}
          g(\gamma) &0 &0 &\gamma/2   \\
           0 &1-2 g(\gamma)&0&0     \\
           0&0&0&0     \\
           \gamma/2&0&0&g(\gamma)     \\
       \end{array}
      \right ]. \ \ \
\end{equation} 
\end{center}
\begin{figure*}
        \centering
                \includegraphics[width=0.6\textwidth]{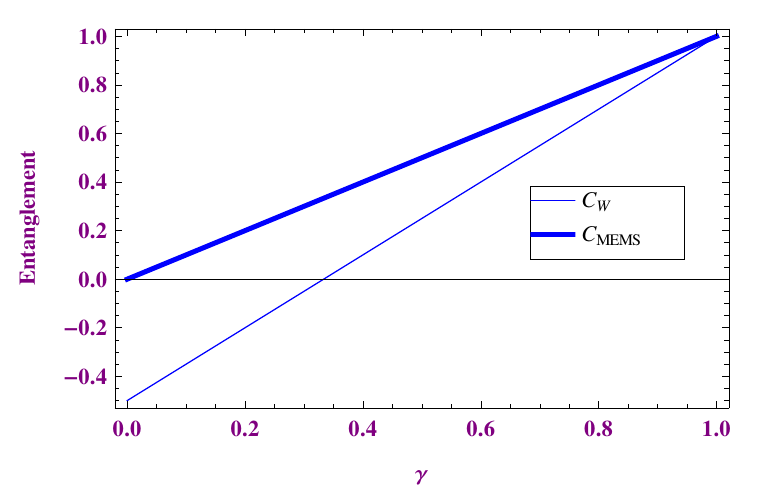}
                \caption{Entanglement plot in Werner states and in MEMS}
\end{figure*}
with $g(\gamma)=\gamma /2$ for $2/3 \leq \gamma\leq 1$ and $g(\gamma)=1/3$ for $0 \leq\gamma < 2/3$. These states carry the concurrence $C(\gamma)=\gamma$. We compare the concurrences in both the states as depicted in Fig. 1. In this figure $C_{W}$ represent the concurrence in Werner states and $C_{MEMS}$ represents the concurrence  in MEMS. The concurrence achieves negative values for Werner states with $\gamma <1/3$, so it is disentangle within the range $0<\gamma <1/3$, while on the other the hand MEMS carry more concurrence than Werner states and hence more entangled than Werner states. 
\section{Reduce density matrix (RDM) and eigenvalue spectrum}
To begin with we consider the auxiliary qubit in pure state as 
\begin{equation}
|\psi\rangle=c_{0}|0\rangle+c_{0}|1\rangle
\end{equation}
which satisfy the normalization condition
\begin{equation}
c_{0}^{2}+c_{1}^{2}=1.
\end{equation}
In order to calculate the RDM first we calculate the time evolution density matrix of the system considered \{i.e Qubit (A), Qubit (B) and auxiliary Qubit (C)\}. It is given as 
\begin{equation}
\rho(t)=U(t)\rho(0)U^{\dagger}(t),
\end{equation}
where U(t) is the unitary time evolution operator obtained in Eq. (6) and $\rho(0)$ is the initial composite density matrix of the open system (formed by qubit A, qubit B and qubit C). First we consider that the two qubits are prepared in Werner states, so the initial density matrix of qubits (A) and qubit (B) pair has to be considered as given in Eq. (9). We obtain the reduced time evolution density matrix by taking the partial transpose operation over the auxiliary qubit C. We observe that the factor $(c_{0}^2+c_{1}^2)$ has been involved with every element of the matrix, so by applying the normalization condition given in Eq. (12), the probability amplitude vanishes from the RDM and do not play any role in reduce dynamics. This result remains same when DM interaction is assumed in any direction ($x$ or $y$ or $z$). We obtain the same result when RDM is calculated for the MEMS. To obtain the RDM for Werner states with DM interaction in x, y and z directions we use the Hamiltonians $H_{x}$, $H_{y}$ and $H_{z}$ given in Eqs. (3)-(5). Further we calculate the eigenvalue spectrum of RDM obtained for Werner states by considering all the Hamiltonians in order to calculate the concurrence by using Eq. (7). We obtain the same eigenvalue spectrum in x, y and z directions. The eigenvalue spectrum is given as 
\begin{equation}
\{\frac{1}{4}(-1+\gamma),\frac{1}{4}(-1+\gamma),\frac{1}{4}(-1+\gamma),\frac{1}{4}(1+3 \gamma)\}.
\end{equation}
\begin{figure*}
        \centering
                \includegraphics[width=\textwidth]{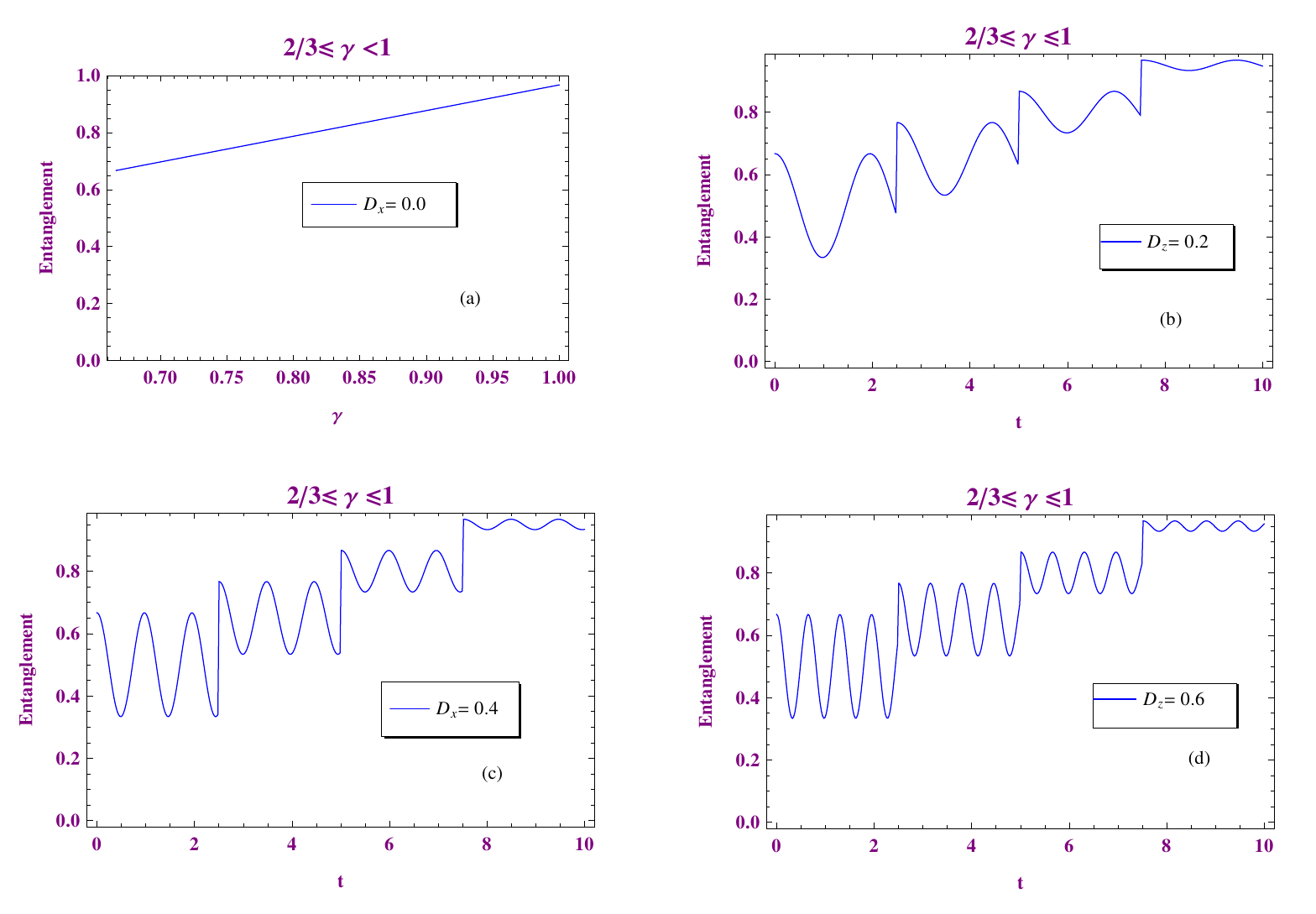}
                \caption{ Entanglement plot vs time  $D_{x}=0.0, $0.2$, 0.4$ and $0.6$}
\end{figure*}
Observing the eigenvalue spectrum we conclude that the components of DM interaction strength, $D_{x}, D_{y}$ and $D_{z}$ in x, y and z directions respectively have not been involved. Hence the Werner states remain unaffected by the x, y and z components of DM interaction. Next we consider another initial state of two qubits (i.e. MEMS). First we consider the case with $2/3\leq\gamma \leq 1$ and later with $0\leq\gamma <2/3$.
\subsection{Case 1: $2/3\leq\gamma \leq 1$}
With the condition $2/3\leq\gamma \leq 1$ for MEMS given in Eq. (10), we obtain the RDM by considering the DM interaction in x, y and z directions and calculate the eigenvalue spectrum $\lambda_{i}$, $i=1,2,3,4,$ for the matrix $(\rho \rho^{f})$ to calculate the concurrence by using Eq. (7), we obtain the eigenvalue spectrum in x direction as   
\begin{equation}
\{0,0,\gamma,(-1+\gamma) \sin[2D_{x}t]^2\}.
\end{equation}
The parameter $D_{x}$ is involved in the spectrum, so the state is affected by the x component of DM interaction. In y direction the eigenvalue spectrum $\lambda_{i}$, $i=1,2,3,4,$ of matrix $(\rho \rho^{f})$ is given as 
\begin{equation}
\{0,0,0,\gamma\}.
\end{equation}
There is no parameter $D_{y}$ involved in the above spectrum, so the state is not being affected by y component of DM interaction. Finally in z direction we obtain the RDM and we get it as a X-structured matrix given below 
\begin{center}
\begin{equation}
\rho=\left[ \begin{array}{cccc}
          \gamma/2 &0 &0 &\gamma/2   \\
           0 &(1-\gamma)\cos[2D_{z}t]^{2}&\frac{1}{2}(-1+\gamma)\sin[4D_{z}t]&0     \\
           0&\frac{1}{2}(-1+\gamma)\sin[4D_{z}t]&(1-\gamma)\sin[2D_{z}t]^{2}&0     \\
           \gamma/2&0&0&\gamma /2    \\
       \end{array}
      \right ]. \ \ \
\end{equation} 
\end{center}
The eigenvalue spectrum $\lambda_{i}$, $i=1,2,3,4,$ of the matrix $(\rho \rho^{f})$ is obtained as 
\begin{equation}
\{0,0,\gamma,(-1+\gamma) \sin[4D_{z}t]^2\}.
\end{equation}
\begin{figure*}
        \centering
                \includegraphics[width=\textwidth]{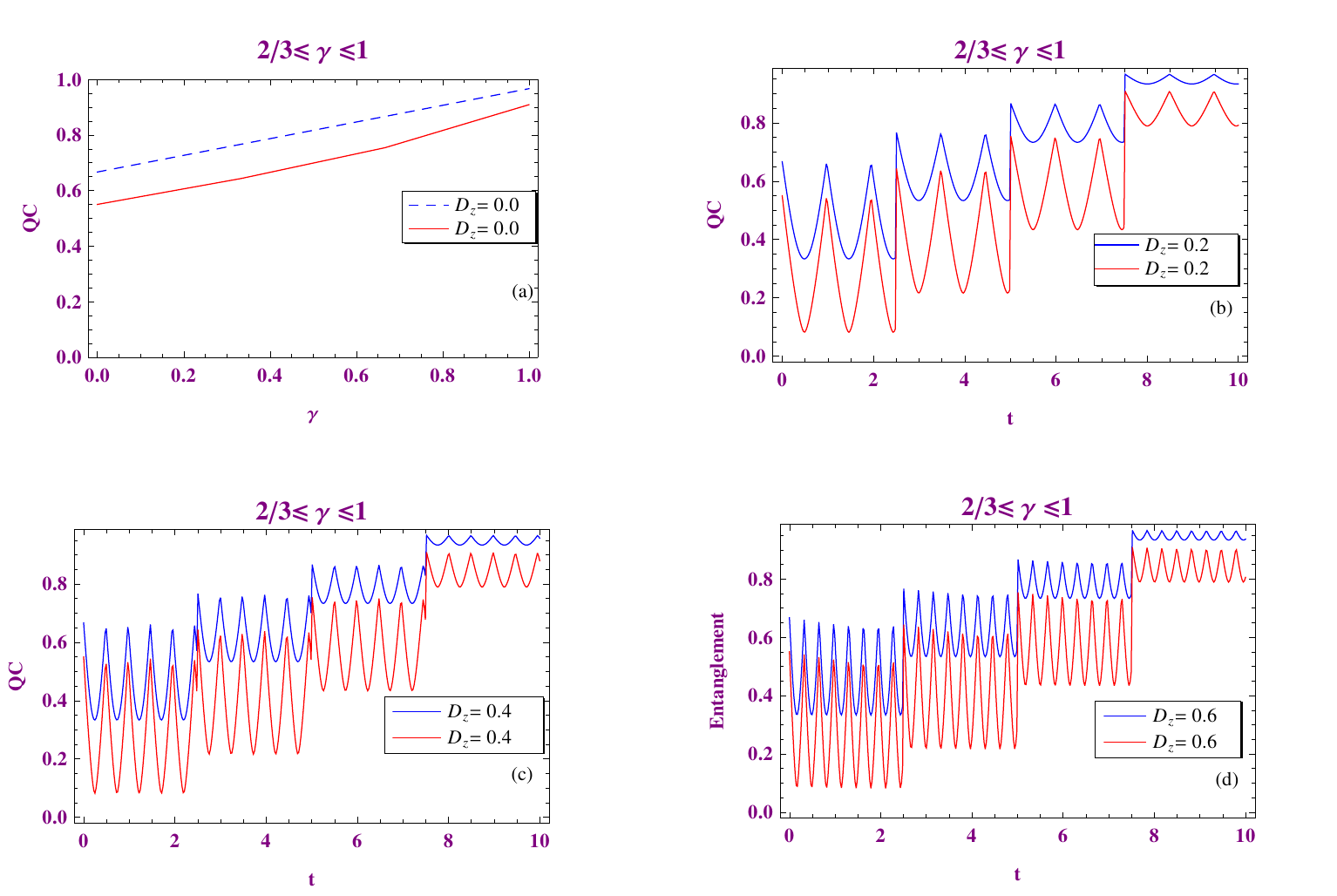}
                \caption{ Variation of quantum correlations, QC (entanglement and quantum discord) for $D_{z}=0.0, 0.2, 0.4$ and $0.6$.}
\end{figure*}
Observing the eigenvalue spectrum we conclude that the state is affected by z component of DM interaction. At this point we also mention that the quantum discord is easy to calculate in the matrix, as it is of X-structure type.
\subsection{Case 2: $0\leq \gamma <2/3$}
Under this case we obtain the RDM by considering the DM interaction in x, y and z directions for MEMS. First we obtain the eigenvalue spectrum $\lambda_{i}$, $i=1,2,3,4,$ of the matrix $(\rho \rho^{f})$ when DM interaction is along the x direction as given below 
\begin{eqnarray}
\{\frac{1}{2},0,\frac{\sqrt{(3-\sqrt{5-4\cos[4D_{x}t]}-2\cos[4D_{x}t])}}{6\sqrt{2}}, \nonumber \\
\frac{\sqrt{(3+\sqrt{5-4\cos[4D_{x}t]}-2\cos[4D_{x}t])}}{6\sqrt{2}}\}.
\end{eqnarray}
We observer that parameter $D_{x}$ is involved in the eigenvalue spectrum and hence the state is affected by the $D_{x}$. Next we calculate the same when DM interaction is assumed along the y direction. We obtain the eigenvalue spectrum $\lambda_{i}$, $i=1,2,3,4,$ of the matrix $(\rho \rho^{f})$ as \\
\begin{equation}
\{\frac{1}{2},\frac{1}{6},0,0\}.
\end{equation} 
\begin{figure*}
        \centering
                \includegraphics[width=\textwidth]{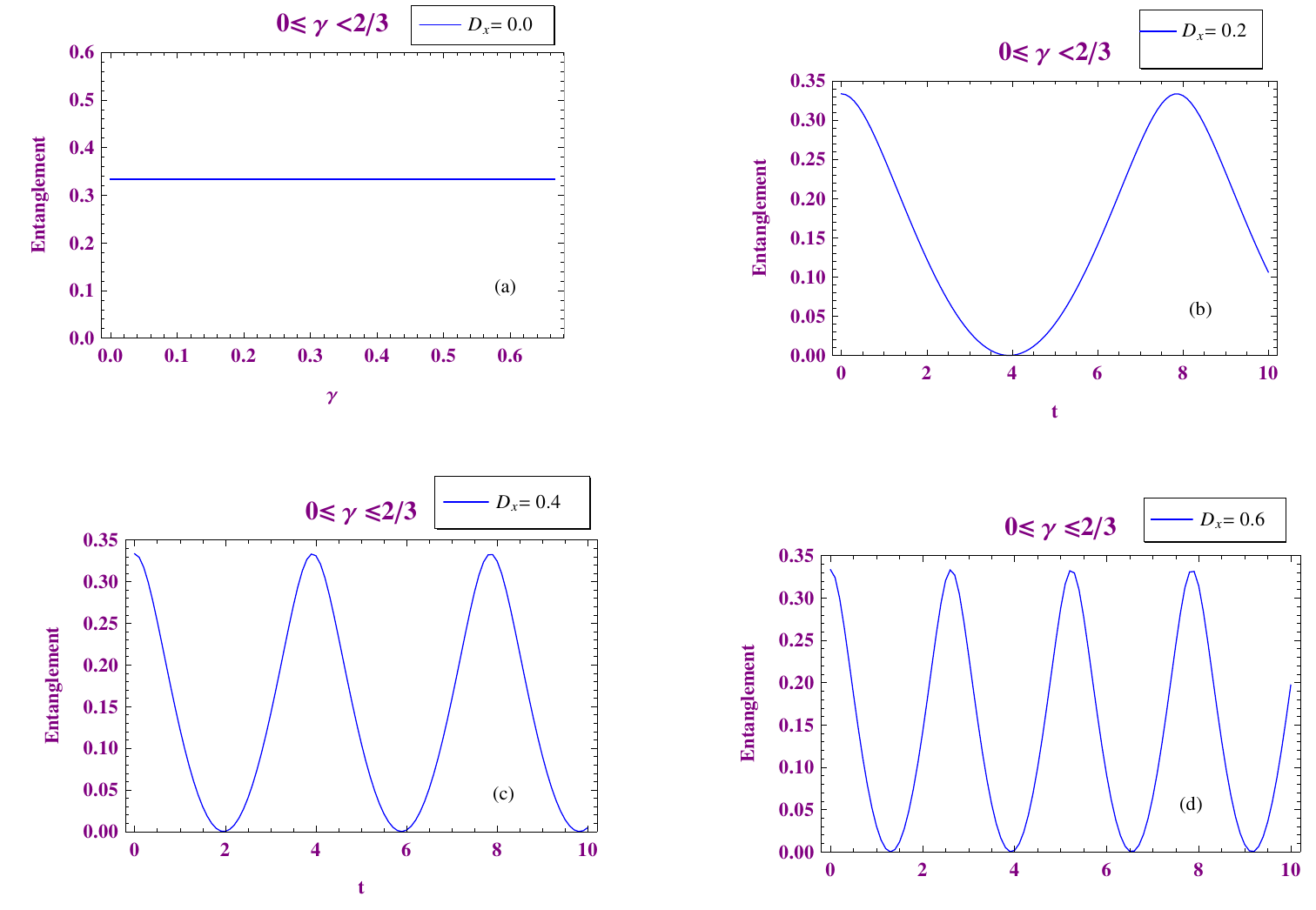}
                \caption{ Entanglement vs. time plot}
\end{figure*}
Here we observe that there is no parameter $D_{y}$ involved, so the state is not affected by the y component of DM interaction. Next we obtain RDM and its eigenvalue spectrum, when DM interaction is assumed along the z direction, we obtain the reduce density matrix as given below  
\begin{center}
\begin{equation}
\rho= \left[ \begin{array}{cccc}
	\frac{1}{3} &0 &0 &\frac{1}{6}   \\
           0 &\frac{1}{3}\cos[2D_{z}t]^{2} &-\frac{1}{6}\sin[4D_{z}t]&0     \\
           0&-\frac{1}{6}\sin[4D_{z}t]&\frac{1}{3}\sin[2D_{z}t]^{2}&0     \\
           	\frac{1}{6}&0&0&\frac{1}{3}    \\
       \end{array}
      \right ]. \ \ \
\end{equation} \label{eq:b3}
\end{center}
The above matrix is X-structured, so quantum discord can be easily calculated. We obtain the eigenvalue spectrum $\lambda_{i}$, $i=1,2,3,4,$ of the matrix $(\rho \rho^{f})$ as follows 
\begin{equation}
\{\frac{1}{2},\frac{1}{6},0,\frac{1}{3}\sin[4D_{z}t]\}.
\end{equation}
Observing the eigenvalue spectrum we conclude that the state is affected by z component of DM interaction.
\begin{figure*}
        \centering
                \includegraphics[width=\textwidth]{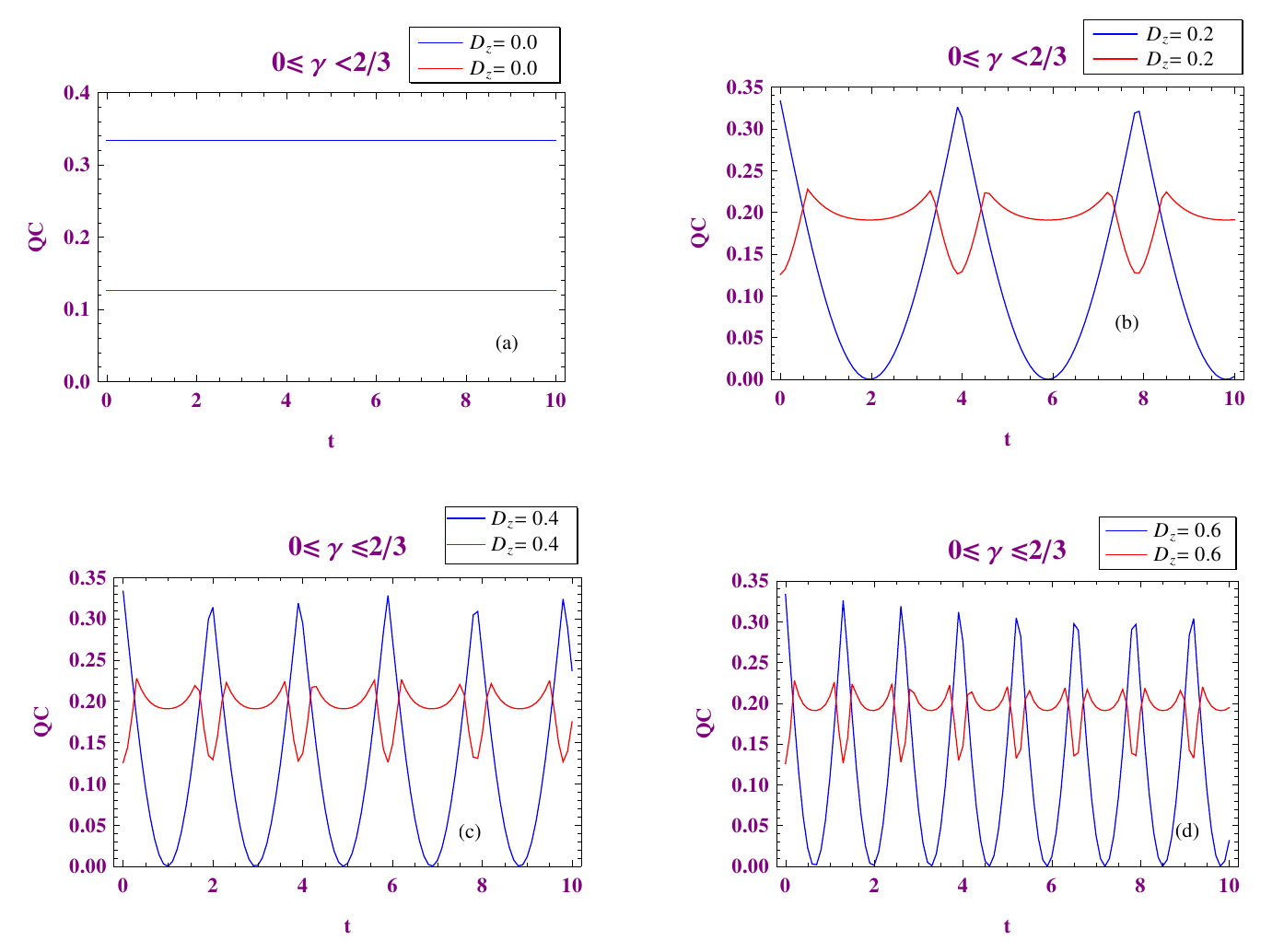}
                \caption{Variation Quantum correlations, QC (entanglement and quantum discord) for $D_{z}=0.0, 0.2, 0.4$ and $0.6$.}
\end{figure*}
\section{Influence of DM interaction on quantum correlations}
In this section we study the influence of $D_{x}$, $D_{y}$ and $D_{z}$ on MEMS for Case 1 and Case 2. We recall that Werner states are not affected by any component of DM interaction. So it is not possible to manipulate the quantum correlations in two qubit Werner states by DM interaction strength and probability amplitude of auxiliary qubit. In this study we are able to calculate the quantum discord for MEMS only when DM interaction is along the z direction, because in this case we obtain that the density matrix is X-structured. Here we reiterate that MEMS are not affected by y-component of DM interaction $(D_{y})$. Throughout this paper we plot the quantum discord by red color and entanglement by blue color.

First we study the quantum correlation for case 1 (i.e. $2/3\leq \gamma \leq 1$) in MEMS for different values of $D_{x}$ and $D_{z}$. For $D_{x}$ we calculate the concurrence (Eq. 7) by using the eigenvalue spectrum given by Eq. (15). We obtain the entanglement in the absence of x-component of DM interaction $(D_{x}=0)$ and the result is shown in Fig. 2(a). We observe that as the value of $\gamma$ increases the entanglement increases linearly and achieve the maximum value 1. Now we increase the value of $D_{x}$ and plot the graphs for $D_{x}=0.2, 0.4$ and $0.6$ in Figs. 2(b), (c) and (d) respectively. We observe that $D_{x}$ excite the entanglement and fluctuate it. Initially at $t=0$, the entanglement achieves the value $0.67$ as seen in figures, but as time advances it is excited and makes oscillations. As the strength of $D_{x}$ increases the frequency of oscillations of entanglement increases. Next we plot the effect of $D_{z}$ on entanglement and quantum discord for the same case. For this purpose we use the X-structured RDM given in Eq. (17). We plot both entanglement and quantum discord with different values of $D_{z}$. First we calculate the quantum correlations in the absence of z-component of DM interaction $(D_{z}=0)$. We plot this result in Fig. 3(a). We conclude that the entanglement achieves the maximum value as $1.0$, while the quantum discord achieves less value $(0.95)$ than entanglement. Further we plot the quantum correlations for higher values of $D_{z}=0.2, 0.4$ and $0.6$ in Figs. 3(b), (c) and (d) respectively. Initially at $t=0$, the entanglement achieves the value $0.67$ but quantum discord achieves the value as $0.55$. As the time advances both entanglement and quantum discord oscillates and quantum discord achieves less value than entanglement. When the strength of DM interaction increases the frequency of oscillations of quantum discord and entanglement increases. With the above results discussed we conclude that the component $D_{x}$ and $D_{z}$ do not produce the ESD in MEMS.

Next we study the effect of $D_{x}$ and $D_{z}$ for the Case 2  (i.e. $0\leq\gamma <2/3$). First we calculate the concurrence (Eq. 7) by using the eigenvalue spectrum given by Eq. (19) for $D_{x}$. We plot the effect of $D_{x}$ on entanglement for $D_{x}=0.0, 0.2, 0.4$ and $0.6$ in Figs. 4(a), (b), (c) and (d) respectively. In Fig. 4(a) we plot the entanglement for $D_{x}=0.0$, in this case the entanglement achieves maximum value as $0.34$. Observing the Fig. 4(b) corresponding to $D_{x}=0.2$ we get the result that at $t=0$ the entanglement achieves the value $0.34$, but as time advances the amplitude of the entanglement oscillates during the time interval $3.5\leq t\leq 4.5$, there is a gap in entanglement evolution (known as ESD). Further at $D_{x}=0.4$ we observe that the oscillations of entanglement increases and the entanglement smoothly goes to zero for a very small interval of time. But at $D_{x}=0.6$, the ESD zone almost vanishes and entanglement tends to zero and rise again. Next we calculate the quantum correlations for $D_{z}$ by using the X-structured matrix given in Eq. (21) and plot in Figs. 5 (a), 5(b), (c) and (d) for $D_{z}=0.0, 0.2, 0.4$ and $0.6$ respectively. In Fig. 5(a) we plot the quantum correlation in the absence of DM interaction $(D_{z}=0)$ and conclude that the entanglement achieves higher value than quantum discord. Further we increase the strength of DM interaction ($D_{z}=0.2$) and plot the quantum correlations in Fig. 5(b). Here we observe that quantum correlations fluctuate. At ($t=0$) the value of quantum discord is $0.12$, while the entanglement value is $0.35$. Here we observe that quantum discord is less than entanglement. Again we find the phenomenon of ESD initially during the time interval of $1.8\leq t\leq 2.1$, but corresponding to this time interval the quantum discord is present. As the time advances the oscillations of both entanglement and quantum discord increases. In Fig. 5(c), with $D_{z}=0.4$ we observe that the entanglement smoothly goes to zero and rise again. But in Fig. 5(d) at $D_{z}=0.6$, we observe that the phenomenon of ESD vanishes and entanglement tends to zero and rise again. Here we mention that in the present case the states go under ESD by DM interaction during the short interval of time. It has also been noted that the frequency of ESD in MEMS for $D_{x}$ is half in comparison to the component $D_{z}$. So the DM interaction in z direction is more intense than x direction.  

\section{Conclusion}
In this paper we have investigated the influence of x, y and z components of DM interaction on quantum correlations of two qubits. The qubits are initially prepared in Werner states and MEMS. We find that two qubit Werner states are not being affected by any component of DM interaction. While on the other hand the MEMS are being affected by the x and z components of DM interaction and remain unaffected by y component of DM interaction. Here we also observe that the state of auxiliary qubit (i.e Probability amplitude) do not play any role in manipulating the quantum correlation in both the Werner and two qubit maximally entanled mixed states. So one can avoid the intention to prepare the state of auxiliary qubit for the same purpose. The avoiding state preparation can help in cost reduction in quantum information processing. We also observe the phenomenon ESD in MEMS in the case 2 $(0\leq \gamma <2/3)$, while there is no ESD with the case 1 $(2/3\leq \gamma \leq 1)$. When $0\leq \gamma <2/3$ the frequency of ESD is half induced by x component of DM interaction in comparison to z component. So the DM interaction is more intense in z direction in comparison to x direction. Further we found that the quantum discord achieves less values than entanglement in MEMS with the case $0\leq \gamma <2/3$.
\appendix
\section{Quantum discord}
To make this paper self-contained we briefly discuss quantum discord here \cite{SLu}. It is known that a bipartite quantum state can carry classical and non-classical correlations. A theoretic measure of information is the mutual information in information theory. The mutual information stored in two classical distributions, equipped with two random variables $X$ and $Y$, assuming values$\{x_{i}\}$ and $\{y_{i}\}$ with probabilities $\{p_{i}\}$ and $\{p_{j}\}$, is defined as 
\begin{equation}
I(X:Y)=H(X)+H(Y)-H(X,Y),
\end{equation} 
\begin{center}
or 
\end{center} 
\begin{equation}
I(X:Y)=H(X)-H(X|Y),
\end{equation}
where $H(X)=-\Sigma_{i}p_{i}log_{2}p_{i}$ and $H(Y)=-\Sigma_{i}q_{i}log_{2}q_{i}$ are the Shannon entropies associated with random variables $X$ and $Y$ respectively, $H(X,Y)$ is the joint entropy of $X$ and $Y$ and $H(X|Y)$ is the conditional entropy of $X$ and $Y$. From classical point of view both the above Eqs. (23) and (24) are equivalent but in quantum mechanical scenario both the expressions are different. Let us consider a bipartite quantum system equipped with parties A and B described by a composite density matrix $\rho^{AB}$with marginal density matrices $\rho^{A}$ and $\rho^{B}$. The quantum analogues of mutual information of Eqs. (23) and (24) is obtained by replacing the Shannon entropies by von Neumann entropies as \begin{equation}
I(\rho^{AB})=S(\rho^{A})+S(\rho^{B})-S(\rho^{AB})
\end{equation}
\begin{center}
and 
\end{center} 
\begin{equation}
I(\rho^{AB})=S(\rho^{A})-S(\rho
^{A}|\rho^{B}),
\end{equation} 
where $S(\rho^{A})=-Tr(\rho^{A}log_{2}\rho^{A})$ and $S(\rho^{B})=-Tr(\rho^{B}log_{2}\rho^{B})$ are the von Neumann entropies of the subsystem A and B respectively and $S(\rho^{AB})=Tr(\rho^{AB})log_{2}\rho^{AB}$ is the joint von Neumann entropy of the composite quantum system AB. The quantity given by Eq.(26) is obscure, because the conditional entropy $S(\rho^{A}|\rho^{B})$ depends
upon measurement performed on either one of the party A or B. So both the Eqs. (25) and (26) are not equivalent from quantum mechanical point of view. The von Neumann projective measurement performed on system A projects the system into a statistical ensemble $\{\rho^{B},p_{k}^{A}\}$, such that 
\begin{equation}
\rho^{AB}\longmapsto \rho_{k}^{B}=\frac{(A_{k}\otimes I_{B})\rho^{AB}(A_{k}\otimes I_{B})}{p_{k}},
\end{equation}
where 
\begin{eqnarray}
p_{k}=Tr[(A_{k}\otimes I_{B})\rho^{AB} (A_{k} \otimes I_{B})] \\
A_{k}=UM_{k}U^{\dagger}, \\
M_{k}=|k\rangle \langle k|, k=0,1 \nonumber
\end{eqnarray} \\
It is known that any unitary can be written up to a constant phase as \\
\begin{equation}
U=tI+y.\sigma,
\end{equation}
with $t\in R$, $y=(y_{1},y_{2},y_{3})\in R^{3}$, $\sigma$ is the Pauli vector of qubit and \\
\begin{center}
 $t^{2}+y_{1}^2+y_{2}^2+y_{3}^2=1$.
\end{center} 
Now Eq. (26) becomes, with respect to to the projective measurement $A_{k}$, as \\
\begin{equation}
I(\rho^{AB}|A_{k}):=S(\rho^{B})-S(\rho^{AB}|A_{k}),
\end{equation} 
where $S(\rho^{AB}|A_{k}):=\Sigma p_{k}S(\rho_{k}^{B})$, is the conditional quantum entropy based on the measurement $A_{k}$. Classical correlations is defined as the maximization of
$I(\rho^{Ab}|A_{k})$ over all possible measurements $A_{k}$ and is given by 
\begin{equation}
C(\rho^{AB}):=sup_{\{A_{k}\}}I(\rho^{AB}|A_{k}).
\end{equation}\\
On simplifying we get 
\begin{equation}
C(\rho^{AB}):=S(\rho^{B})-\min_{\{A_{k}\}}S(\rho^{B}|A_{k}).
\end{equation}\\
Quantum discord is given by the difference of mutual quantum information $I(\rho^{AB})$ given in Eq. (25) and classical correlations $C(\rho^{AB})$ given in Eq. (33). So we obtained the expression of quantum discord as 
\begin{equation}
Q(\rho^{AB})=I(\rho^{AB})-C(\rho^{AB}) \nonumber
=S(\rho^{A})-S(\rho^{AB})+\min_{\{{A_{k}}\}}S(\rho^{B}|A_{k}).
\end{equation}
The calculation of quantum discord is based on complex maximization procedure and obtaining an analytic expression of $C(\rho_{AB})$ for general states is not an
easy task. Here we write our x-structured density matrix as follows
\begin{center}
\begin{equation}
\rho^{AB}= \left[ \begin{array}{cccc}
          \rho_{11} &0 &0 &\rho_{14}    \\
           0 &\rho_{22}&\rho_{23}&0     \\
           0&\rho_{32}&\rho_{33}&0     \\
           \rho_{41}&0&0&\rho_{44}    \\
       \end{array}
      \right ]. \ \ \
\end{equation} 
\end{center}
Here we use the method to calculate the quantities $Q(\rho^{AB})$, $CC(\rho^{AB})$ given by C. Z. Wang et al.\cite{qd}, So we obtain $Q(\rho^{AB})$, $CC(\rho^{AB})$ as follows \\
\begin{eqnarray}
CC(\rho^{AB})=max(CC_{1},CC_{2}),\\
QD(\rho^{AB})=min(QD_{1},QD_{2}),
\end{eqnarray}
where
\begin{center}
$CC_{j}=H(\rho_{11}+\rho_{22})-D_{j}$, \\
$QD_{j}=H(\rho_{11}+\rho_{33})+\Sigma_{k=1}^{4}\lambda_{k}Log_{2}\lambda_{k}+D_{j}$,
\end{center} 
and \\
\begin{center}
$D_{1}=H(\tau)$, $D_{2}=-\Sigma_{j=1}^{4}\rho_{jj}log_{2}\rho_{jj}-H(\rho_{11}+\rho_{33})$,
\end{center}
\begin{center}
$H(\rho_{11}+\rho_{22})=-(\rho_{11}+\rho_{22})log_{2}(\rho_{11}+\rho_{22})-[1-(\rho_{11}+\rho_{22})]log_{2}[1-(\rho_{11}+\rho_{22})]$,
\end{center}
\begin{center}
$\tau=\frac{1+\sqrt{[1-2(\rho_{33}+\rho_{44})]^{2}+4(|\rho_{14}|+|\rho_{23}|)^{2}}}{2}$,
\end{center}
where $\lambda_{k}$ are the eigenvalues of $\rho^{AB}$ and $\rho_{33}$, $\rho_{44}$, $\rho_{14}$ and $\rho_{23}$ are the elements of x-structured density matrix given in Eq. (35).\\ 
\section{Mathematica code}
Under this section we provide the mathematica code to calculate the quantum discord for X-structured density matrices. The code is given below. \\ \\
$m=\{\{\rho_{11},0,0,\rho_{14}\},\{0,\rho_{22},\rho_{23},0\},\{ 0,\rho_{32},\rho_{33},0\},\{\rho_{14},0,0,\rho_{44}\}\};$\\
$e$=Eigenvalues [m];\\
$\lambda_{1}=e[[1]];$ $\lambda_{2}=e[[2]];$ $\lambda_{3}=e[[3]];$ $\lambda_{4}=e[[4]];$ \\
$H[x_{-}]:=$\\
Module[$\{ x1=x\},$ $-x1$ Log2[$x1$]-$(1-x1)$ Log2[$1-x1$]]\\
Log2[$x_{-}$]=If[$x=0$,0,Log2[x]]; \\
p=FullSimplify[(ComplexExpand [Abs$[m[[1]][[4]]]$+ComplexExpand [Abs$[m[[2]][[3]]])^{2}$]; \\
$\tau=(1/2)*(1+\sqrt((1-2(m[[3]][[3]]+m[[4]][[4]]))^2+4(p)))$;\\
DD1=N[H[$\tau$]];\\
a=-m[[1]][[1]]Log2[m[[1]]m[[1]]];\\
b=-m[[2]][[2]] Log2[m[[2]][[2]]]; \\
c=-m[[3]][[3]] Log2 [m[[3]][[3]]]; \\
d=-m[[4]][[4]] Log2 [m[[4]][[4]]]; \\
f=-H[m[[1]][[1]]]+m[[3]][[3]]; \\
DD2=a+b+c+d+f; \\
Q11=H[m[[1]][[1]]+m[[3]][[3]]]+$\lambda_{1}$ Log2 [$\lambda_{1}$]+$\lambda_{2}$ Log2 [$\lambda_{2}$]+$\lambda_{3}$ Log2 [$\lambda_{3}$]+$\lambda_{4}$ Log2 [$\lambda_{4}$]+DD1;\\
Q22=H[m[[1]][[1]]+m[[3]][[3]]]+$\lambda_{1}$ Log2 [$\lambda_{1}$]+$\lambda_{2}$ Log2 [$\lambda_{2}$]+$\lambda_{3}$ Log2 [$\lambda_{3}$]+$\lambda_{4}$ Log2 [$\lambda_{4}$]+DD2; \\
QD=N[Min[Q11,Q22]]; \\
Quit[]

\end{document}